\documentclass[12pt]{article}
\usepackage{graphicx}
\usepackage{amssymb}

\begin{document}
\title{\bf Associative Memory by Recurrent Neural Networks 
with Delay Elements
\thanks{This research was partially supported by the Ministry of Education, 
Science, Sports and Culture, Japan, Grant-in-Aid for Scientific Research,
13780303.}} 

\author{
Seiji Miyoshi
$\hspace{-1.5mm}$\thanks{Corresponding author. 
E-mail:miyoshi@kobe-kosen.ac.jp} , 
Hiro-Fumi Yanai${}^{\dagger\dagger}$,
and Masato Okada${}^{\dagger\dagger\dagger , \dagger\dagger\dagger\dagger}$\\
\\
\small
${}^{\dagger}$Department of Electronic Engineering, 
Kobe City College of Technology\\
\small
8-3 Gakuen-Higashimachi, Nishi-ku, Kobe 651-2194, Japan\\
\\
\small
${}^{\dagger\dagger}$Department of Media and Telecommunications,
Faculty of Engineering,
Ibaraki University\\
\small
Naka-Narusawa, Hitachi, Ibaraki 316-8511, Japan\\
\\
\small
${}^{\dagger\dagger\dagger}$Exploratory Research for Advanced Technology, 
Japan Science and Technology\\
\small
2-2 Hikari-dai, Seika-cho, Soraku-gun, Kyoto 619-0288, Japan\\
\\
\small
${}^{\dagger\dagger\dagger\dagger}$Laboratory for
Mathematical Neuroscience, RIKEN Brain Science Institute\\
\small
2-1 Hirosawa, Wako, Saitama 351-0198, Japan
}
\date{}

\maketitle

\noindent \mbox{}\hrulefill
\begin{flushleft}
{\bf Abstract}
\end{flushleft}

The synapses of real neural systems seem to have delays.
Therefore, it is worthwhile
to analyze associative memory models with delayed synapses.
Thus, a sequential associative memory model with delayed synapses is
discussed, where a discrete synchronous updating rule and
a correlation learning rule are employed.
Its dynamic properties are analyzed
by the statistical neurodynamics.
In this paper, 
we first re-derive the Yanai-Kim theory, 
which involves macrodynamical equations
for the dynamics of the network with serial delay elements.
Since 
their theory needs a computational complexity of ${\cal O}(L^4 t)$
to obtain the macroscopic state at time step $t$
where $L$ is the length of delay,
it is intractable to discuss the macroscopic properties
for a large $L$ limit.
Thus, we derive steady state equations
using the discrete Fourier transformation,
where the computational 
complexity does not formally depend on $L$.
We show that
the storage capacity $\alpha_C$ is in proportion to the delay length
$L$ with a large $L$ limit,
and the proportion constant is 0.195, i.e., $\alpha_C = 0.195 L$.
These results are supported by computer simulations.

\vspace{10mm}
\noindent
{\it Key words}: sequential associative memory, neural network, delay, 
statistical neurodynamics

\noindent \mbox{}\hrulefill

\section{Introduction}
Associative memory models of neural network can be mainly classified
into two types (Okada, 1996; Fukushima, 1973; Miyoshi \& Okada, 2000).
The first is the auto associative memory model
where memory patterns are stored as equilibrium states of the network. 
The second type is the sequence processing model, which stores 
the sequence of memory patterns.

As a learning algorithm for storing memory patterns,
the correlation learning algorithm
based on Hebb's rule is well known. 
The storage capacity,
that is, how many memory patterns can be stably stored
with respect to the number of neurons,
is the one of the most important properties of associative memory
models.
Hopfield
showed that the storage capacity 
of the auto associative memory model
using correlation learning is about 0.15 by computer
simulation (Hopfield, 1982).
On the other hand,
many theoretical analyses have been done on the correlation
learning type associative memory model
(Okada, 1996).
As typical analytical methods, there are the replica method
(Sherrington \& Kirkpatrick, 1975; Amit et al., 1985a; Amit et al., 1985b)
and the SCSNA
(Shiino \& Fukai, 1992)
for the equilibrium state of the auto associative memory models.
There is also the statistical neurodynamics
(Amari \& Maginu, 1988)
for handling retrieval dynamics.
By these theories,
it became clear that 
the storage capacity of the auto-associative memory model is 0.138 and
that of the sequence processing model is 0.269
(Amari, 1988; D\"{u}ring et al., 1998; Kawamura \& Okada, 2002).
Furthermore,
it has become well-known that
the analysis of the dynamics for the auto-associative model
is more difficult than that for the sequence processing model.
However, Okada
(Okada, 1995) succeeded in explaining the dynamics of
the retrieval process quantitatively by extending the statistical
neurodynamics
(Amari \& Maginu, 1988).

On the other hand, the synapses of real neural systems
seem to have delays.
Therefore, it is very important to analyze the associative memory model
with delayed synapses.
Computer simulation is a powerful method for investigating
the properties of the neural network.
However, there is a limit on the number of neurons.
In particular,
computer simulation for a network
that has large delay steps is realistically impossible
considering the required amount of calculation and memory.
Therefore,
the theoretical and analytical approach is indispensable
to research on delayed networks.

A neural network in which each neuron has delay elements
(Fukushima, 1973; Yanai \& Kim, 1993; Miyoshi \& Nakayama, 1995)
was analyzed by Yanai and Kim
(Yanai \& Kim, 1993).
They analyzed the delayed network
using the statistical neurodynamics
(Amari \& Maginu, 1988; Amari, 1988)
and
derived macrodynamical equations for the dynamics.
Their theory closely agrees with
the results of computer simulation. 

In this paper, after defining the model, 
the Yanai-Kim theory is re-derived by
using the statistical neurodynamics
(Miyoshi \& Okada, 2002).
We show that their macrodynamical equations make clear
that the dynamics of the network and the phase transition points change
with the initial conditions.
The Yanai-Kim theory needs a computational complexity of ${\cal O}(L^4 t)$
to obtain the macrodynamics,
where $L$ and $t$ are the length of delay and the time step,
respectively.
This means that it is indispensable
to discuss the macroscopic properties
for a large $L$ limit.
Thus, we derive the macroscopic steady state equations 
by using the discrete Fourier transformation.
Using the derived steady state equations, 
the storage capacity can be quantitatively discussed
even for a large $L$ limit.
Then, it
becomes clear that the phase transition
point calculated from the macroscopic steady state equations 
agrees with the phase transition point obtained by time-dependent 
calculation with a sufficient number of time steps
from the optimum initial condition.
Furthermore, it becomes clear that in the case of 
large delay length $L$, the storage capacity is in proportion to 
the length and the proportion constant is 0.195.
These results are supported by computer simulation.

\section{Model of delayed network}
\begin{figure*}[t]
\begin{center}
\includegraphics[width=0.90\linewidth,keepaspectratio]{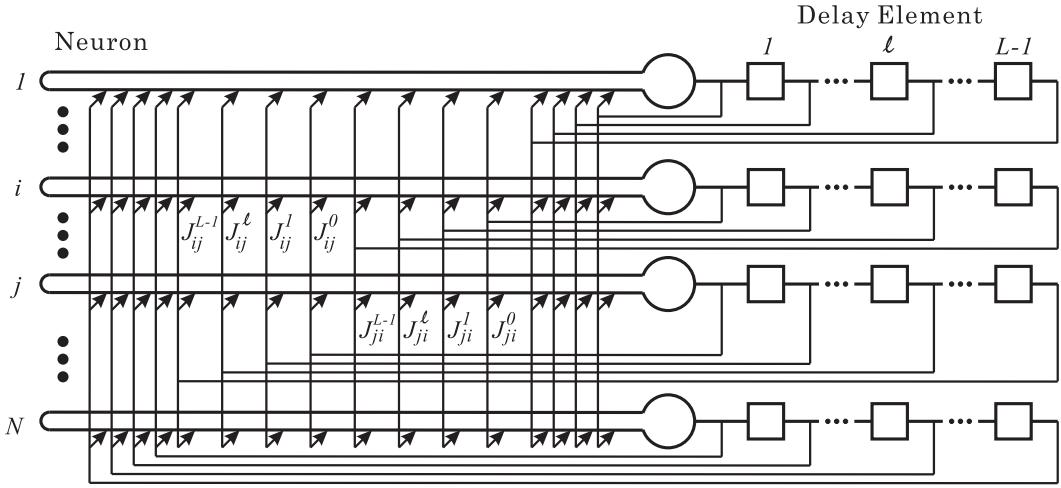}
\caption{Structure of delayed network.}
\label{fig:str}
\end{center}
\end{figure*}

The structure of the delayed network discussed in this paper is 
shown in Figure \ref{fig:str}.
The network has $N$ neurons, and $L-1$ serial delay elements are connected
to each neuron. All neurons as well as all delay elements have synaptic
connections with all neurons.
In this neural network, all neurons and all
delay elements change their states simultaneously.
That is, this network employs a discrete synchronous updating rule.
The output of each neuron is determined as
\begin{eqnarray}
x_i^{t+1} &=& F\left(u_i^t\right) , \label{eqn:xF} \\
u_i^t &=& \sum_{l=0}^{L-1}\sum_{j=1}^N J_{ij}^l x_j^{t-l} ,
\label{eqn:uit}
\end{eqnarray}
where $x_i^t$ denotes 
the output of the $i$th neuron at time $t$, and $J_{ij}^l$
denotes the connection weight from the $l$th delay elements of the $j$th
neuron to the $i$th neuron.
$F\left(\cdot\right)$ is the sign function defined as
\begin{equation}
F\left(u\right)=\mbox{sgn}\left(u\right)=\left\{
\begin{array}{ll}
+1,            & u \geq 0 .  \\
-1,            & u <    0 .
\end{array}
\right.
\label{eqn:sgn}
\end{equation}

Now, let's consider how to store the sequence of $\alpha N$ memory patterns,
$\mbox{\boldmath $\xi$}^1 \rightarrow 
\mbox{\boldmath $\xi$}^2
\rightarrow \cdots \rightarrow \mbox{\boldmath $\xi$}^\mu
 \rightarrow \cdots \rightarrow \mbox{\boldmath $\xi$}^{\alpha N}$.
Here, $\alpha$ and $\alpha N$ are the loading rate
and the length of the sequence, respectively.
Each component $\mbox{\boldmath $\xi$}^{\mu}$ is assumed to be an
independent random variable
that takes a value of either $+1$ or $-1$ according to
the following probabilities,
\begin{equation}
 \mbox{Prob}\left[\xi_i^{\mu}=\pm1\right]=\frac{1}{2} .
\end{equation}

We adopt the following learning method using the correlation learning,
\begin{equation}
J_{ij}^l=\frac{c_l}{N}\sum_\mu \xi_i^{\mu+1+l}\xi_j^\mu ,
\label{eqn:correlation}
\end{equation}
where $c_l$ is the strength of the $l$th delay step.

Correlation learning is an algorithm based 
on Hebb's rule. It is inferior to the error correcting learning
in terms of storage capacity. However, as seen
 from eqn (\ref{eqn:correlation}), 
when adding new patterns, it is not necessary to again learn all patterns
that were
stored in the past. Furthermore, correlation learning 
has been analyzed by many researchers due to its simplicity.

\section{Dynamical behaviors of macroscopic order parameters by 
statistical neurodynamics and discussion} \label{sec:dynamical}
As mentioned above,
a neural network in which each neuron has delay elements
(Fukushima, 1973; Yanai \& Kim, 1993; Miyoshi \& Nakayama, 1995)
was analyzed by Yanai and Kim
(Yanai \& Kim, 1993).
In this section, 
the Yanai-Kim theory is re-derived
by using the statistical neurodynamics
(Miyoshi \& Okada, 2002).
Using the re-derived theory, we discuss the dynamical behaviors
of a delayed network. 
Furthermore,
the relationship between the phase transition point 
and the initial conditions of the delayed network
is discussed.

In the case of a small loading rate $\alpha$, 
if a state close to one or some of the patterns 
stored as a sequence
are given to the network,
the stored sequence of memory patterns is retrieved.
However, when loading rate $\alpha$ increases, 
the memory is broken at a certain $\alpha$. That is, 
even if a state close to one or some of the patterns 
stored as a sequence
is given to the network,
the state of the network tends to leave the stored 
sequence of memory patterns.
Moreover, even if one or some of the patterns themselves
are given to the network,
the state of the network tends to leave
the stored sequence of memory patterns.
This phenomenon, that is, the memory suddenly
becoming unstable at a critical
loading rate, can be considered a kind of phase transition.

We define the overlaps or direction cosine between 
a state $\mbox{\boldmath $x$}^t=\left(x_i^t\right)$ 
appearing in a recall process at time $t$ 
and 
an embedded pattern $\mbox{\boldmath $\xi$}^\mu
=\left(\xi_i^\mu\right)$ as
\begin{equation}
m_\mu^t=\frac{1}{N}\sum_{i=1}^N \xi_i^{\mu} x_i^t.
\label{eqn:mmut}
\end{equation}

By using this definition, when the state of the network at time $t$ and
the $\mu$th pattern agree perfectly, the overlap $m_t^{\mu}$ is equal to unity.
When they have no
correlation, the overlap $m_t^{\mu}$ is equal to zero.
Therefore, the overlap provides a way to measure recall quality.

Amari and Maginu
(Amari \& Maginu, 1988)
proposed the statistical neurodynamics.
This analytical method handles the 
dynamical behavior of the recurrent neural network macroscopically,
where cross-talk noise is regarded as a Gaussian random
variable with a mean of zero and a time-dependent
variance of $\sigma_t^2$. They then derived recursive relations
for the variance and the overlap.

Using eqns (\ref{eqn:uit}), (\ref{eqn:correlation}) and (\ref{eqn:mmut}),
the total input of the $i$th neuron at time $t$ is given as
\begin{eqnarray}
u_i^t
 &=& \sum_{l=0}^{L-1}\sum_{j=1}^N J_{ij}^lx_j^{t-l} \\
 &=& s^t \xi_i^{t+1}+z_i^t \label{eqn:sz}, \\
s^t
 &=& \sum_{l=0}^{L-1} c_l m_{t-l}^{t-l}, \\
z_i^t
 &=& \sum_{l=0}^{L-1}c_l\sum_{\nu \neq t}
     \xi_i^{\nu+1}m_{\nu-l}^{t-l}. \label{eqn:zit}
\end{eqnarray}

The first term in eqn (\ref{eqn:sz}) is the signal that is useful
for recall, while the second term is cross-talk noise which 
prevents $\xi_i^{t+1}$ from being recalled.
This procedure is called a signal-to-noise analysis.

The overlap $m_\mu^t$ can be expressed as
\begin{eqnarray}
m_\mu^t
&=& \frac{1}{N}\sum_{i=1}^N \xi_i^\mu x_i^t \\
&=& \bar{m}_\mu^t + U_t\sum_{l'=0}^{L-1}c_{l'}m_{\mu-l'-1}^{t-l'-1},
\label{eqn:mmut2} \\
\bar{m}_\mu^t
&=& \frac{1}{N}\sum_{i=1}^N \xi_i^\mu F\left(\sum_{l=0}^{L-1}
    \sum_{j=1}^N\frac{c_l}{N}\right. \nonumber \\
& & \hspace{10mm} \times \left. 
    \sum_{\nu \neq \mu-l-1}\xi_i^{\nu+1+l}\xi_j^\nu x_j^{t-l-1}\right),
    \\
U_t
&=& \frac{1}{N}\sum_{i=1}^N 
    F'\left(\sum_{l=0}^{L-1}\sum_{j=1}^N\frac{c_l}{N}\right. 
    \nonumber \\
& & \hspace{10mm} \times \left.\sum_{\nu \neq \mu-l-1}\xi_i^{\nu+1+l}
    \xi_j^\nu x_j^{t-l-1}\right). \label{eqn:Ut}
\end{eqnarray}

Taking into account
the correlation in the cross-talk noise
$z_i^t$, we have derived the following macrodynamical 
equations using eqns (\ref{eqn:xF})-(\ref{eqn:Ut})
(see Appendix).
\begin{eqnarray}
\sigma_t^2
 &=& \sum_{l=0}^{L-1}\sum_{l'=0}^{L-1}c_lc_{l'}v_{t-l,t-l'} ,
 \label{eqn:sigmat2D} \\
v_{t-l,t-l'}
 &=& \alpha \delta_{l,l'} \nonumber \\
 &+& U_{t-l}U_{t-l'} \nonumber \\
 & & \times 
    \sum_{k=0}^{L-1}\sum_{k'=0}^{L-1}c_k c_{k'}v_{t-l-k-1,t-l'-k'-1}
    \nonumber \\
 &+& \alpha \left(c_{l-l'-1}U_{t-l'} + c_{l'-l-1}U_{t-l}\right) ,
 \label{eqn:vD} \\
U_t &=& \sqrt{\frac{2}{\pi}} \frac{1}{\sigma_{t-1}}
     \exp\left(-\frac{\left(s^{t-1}\right)^2}{2\sigma_{t-1}^2}\right) ,
     \label{eqn:UtD} \\
s^t &=& \sum_{l=0}^{L-1} c_l m_{t-l} , \label{eqn:stD} \\
m_{t+1} &=& \mbox{erf}\left(\frac{s^t}{\sqrt{2}\sigma_t}\right) ,
\label{eqn:mt1D}
\end{eqnarray}
where $m_{t}$ denotes $m_{t}^{t}$.
If $t<0$, $m_t=0$ and $U_t=0$.
If $k<0$, $c_k=0$.
If either $k<0$ or $k'<0$, $v_{k,k'}=0$.
The expression
${\rm erf}\left(x\right)\equiv \frac{2}{\sqrt{\pi}}\int_0^x 
\exp\left(-u^2\right)du$
denotes the error function.

Various cases can be considered as the initial condition of 
the network. For example, in the case of {\it the one step set initial
condition}, 
only the states of neurons are set explicitly and 
those of delay elements are all zero. Then $m_0=m_{{\rm init}}$ is 
the only initial condition that is set explicitly and $v_{0,0}=\alpha$.
On the other hand, as one more extreme case, 
{\it the all steps set initial condition} can also be considered,
where the states of all neurons and all delay elements are set to
be close to
the stored pattern sequences.
In this case, $m_l=m_{{\rm init}},\ l=0,\cdots,L-1$ and 
$v_{l,l}=\alpha,\ l=0,\cdots,L-1$.
In the case where all neurons and all delay elements 
are set to $m_{{\rm init}}=1$,
that is, $m_l=1,\ l=0,\cdots,L-1$,
the maximum information for recalling the stored sequence 
of memory patterns is given. 
Therefore, this condition can be called 
{\it the optimum initial condition}.
Furthermore, the storage capacity $\alpha_C$ is defined as
the critical loading rate where
recalling becomes unstable under the optimum initial condition.
We note that the derived macrodynamical equations 
and the Yanai-Kim theory
(Yanai \& Kim, 1993) coincide with each other.

\begin{figure}[htbp]
\begin{center}
\includegraphics[width=0.70\linewidth,keepaspectratio]{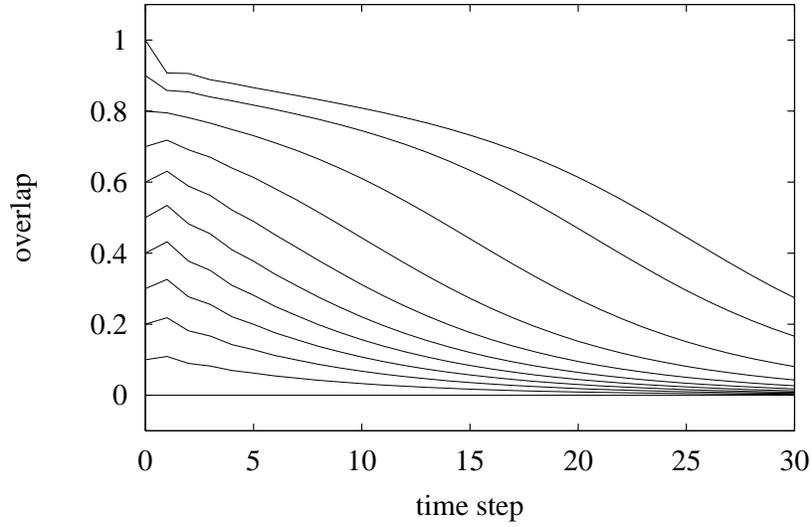}
\caption{Dynamical behaviors of recall process 
($L=2,\alpha=0.5$: theory).}
\label{fig:L2A05t}
\end{center}
\end{figure}

\begin{figure}[htbp]
\begin{center}
\includegraphics[width=0.70\linewidth,keepaspectratio]{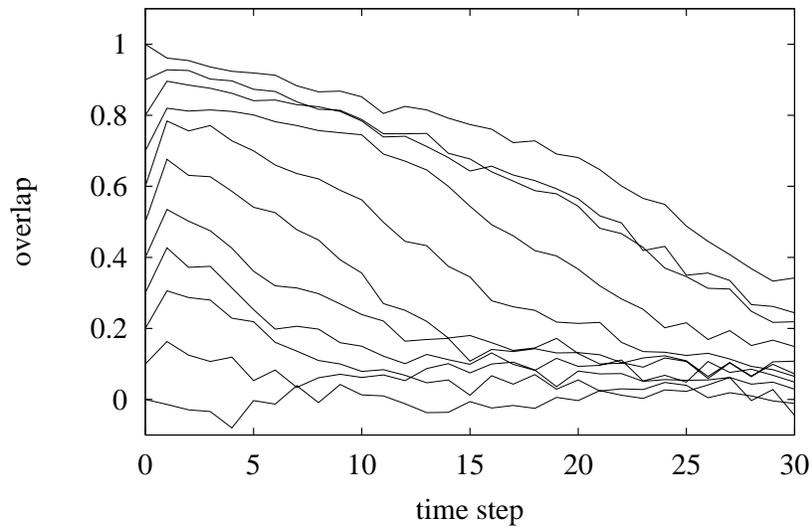}
\caption{Dynamical behaviors of recall process 
($L=2,\alpha=0.5$: computer simulation with $N=2000$).}
\label{fig:L2A05s}
\end{center}
\end{figure}

\begin{figure}[htbp]
\begin{center}
\includegraphics[width=0.70\linewidth,keepaspectratio]{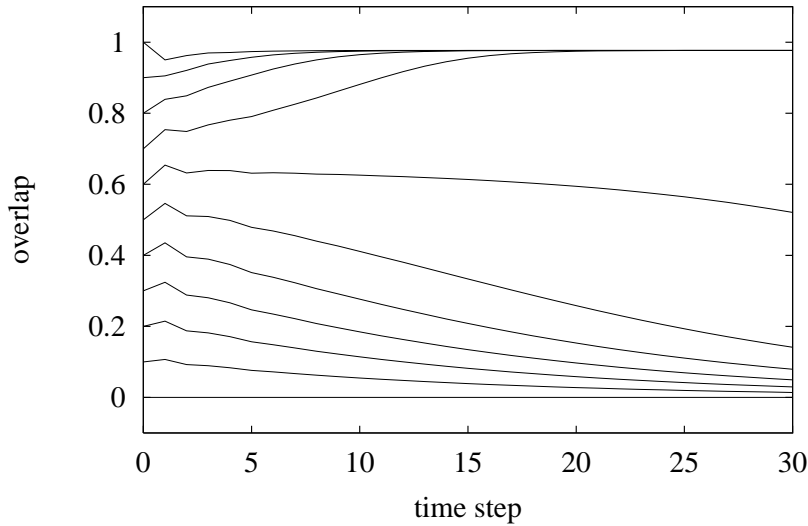}
\caption{Dynamical behaviors of recall process 
($L=3,\alpha=0.5$: theory).}
\label{fig:L3A05t}
\end{center}
\end{figure}

\begin{figure}[htbp]
\begin{center}
\includegraphics[width=0.70\linewidth,keepaspectratio]{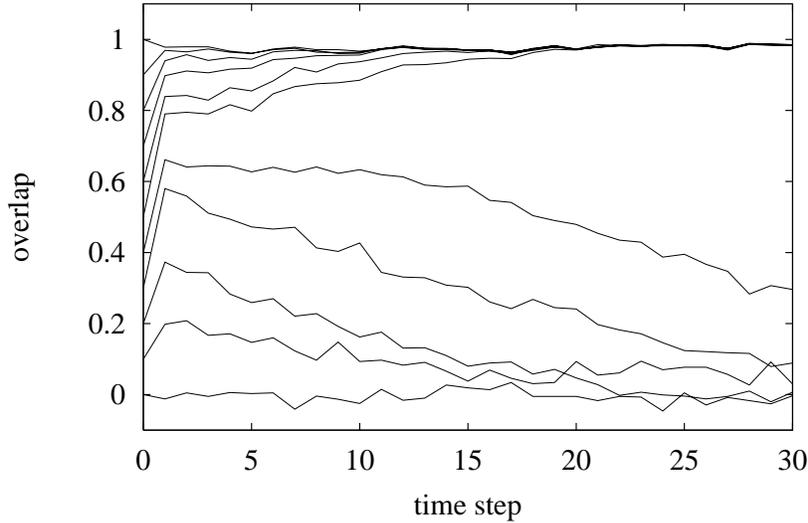}
\caption{Dynamical behaviors of recall process 
($L=3,\alpha=0.5$: computer simulation with $N=2000$).}
\label{fig:L3A05s}
\end{center}
\end{figure}

Some examples of the dynamical behaviors of recall processes by
the above theory and computer simulations are shown in 
Figures \ref{fig:L2A05t}-\ref{fig:L3A05s}. In these figures, 
the abscissa is time $t$, and the ordinate is the overlap $m_t$.
These are the results when the all steps set initial conditions
are given.
Figures \ref{fig:L2A05t} and \ref{fig:L3A05t} are the results 
of 30-step time-dependent theoretical calculation 
by eqns (\ref{eqn:sigmat2D})-(\ref{eqn:mt1D}) using
various initial overlaps.
The details of the computer simulations shown in
Figures \ref{fig:L2A05s} and \ref{fig:L3A05s}
are as follows. 
First, a sequence of random patterns
are generated. The length of the sequence is 
$\alpha N=1000$. Each pattern is a vector in which the 
dimension is $N=2000$. Therefore, the loading rate is $\alpha=0.5$.
Each element of the pattern vectors 
takes a value of either $+1$ or $-1$ with probability $\frac{1}{2}$.
Next, the sequence is stored by
correlation learning in two networks of which the number of 
neurons is $N=2000$
and the lengths of delay $L$ are two and three. 
Then, various pattern sequences are generated
where the initial overlaps with
the 1st--$L$th patterns stored are $0.0-1.0$.
These are given as the initial state of the network.
After that, 30-step calculation is carried out by using 
the updating rule of the network, that is, 
eqns (\ref{eqn:xF})-(\ref{eqn:sgn}).

According to these figures,
the dynamical behaviors of the overlaps obtained by the re-derived theory
are in good agreement with those obtained by 
computer simulation.
Figure \ref{fig:L2A05t} shows the dynamical behaviors
of the overlaps in the case of $L=2$.
When the initial overlaps are $0.1-0.7$, 
the overlaps somewhat increase at the first time step.
However, the recall processes eventually fail
regardless of the initial overlaps.
This fact indicates that the storage capacity $\alpha_C$ in the case of 
$L=2$ is smaller than 0.5. 
On the other hand, Figure \ref{fig:L3A05t} shows 
the dynamical behaviors
of the overlaps in the case of $L=3$.
When the initial overlap is 1.0,
at the first time step
the overlap somewhat decreases. However, then the overlap 
tends to a value near 1.0. 
This fact indicates that the storage capacity $\alpha_C$ in the case of 
$L=3$ is larger than 0.5. 

\begin{figure}[htbp]
\begin{center}
\includegraphics[width=0.70\linewidth,keepaspectratio]{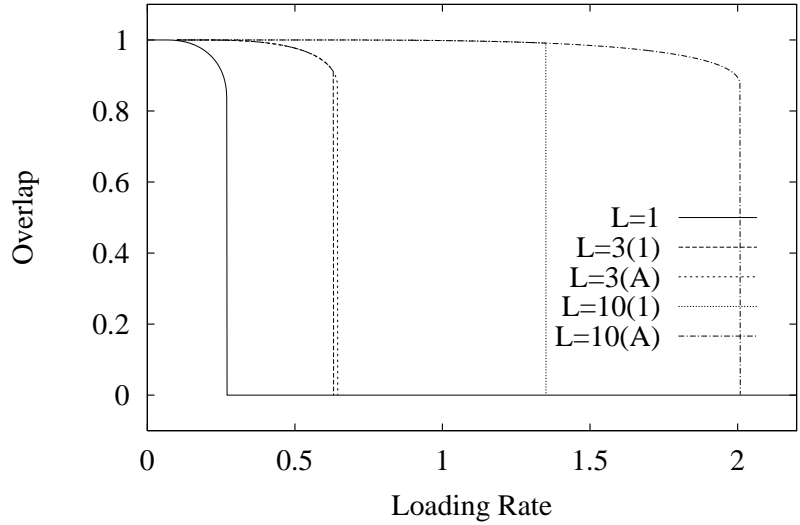}
\caption{Relationship between 
loading rate $\alpha$ and overlap $m$.
``(1)" and ``(A)" indicate one step set initial condition
 and all steps set initial condition, respectively (theory).}
\label{fig:Ld1_At}
\end{center}
\end{figure}

\begin{figure}[htbp]
\begin{center}
\includegraphics[width=0.70\linewidth,keepaspectratio]{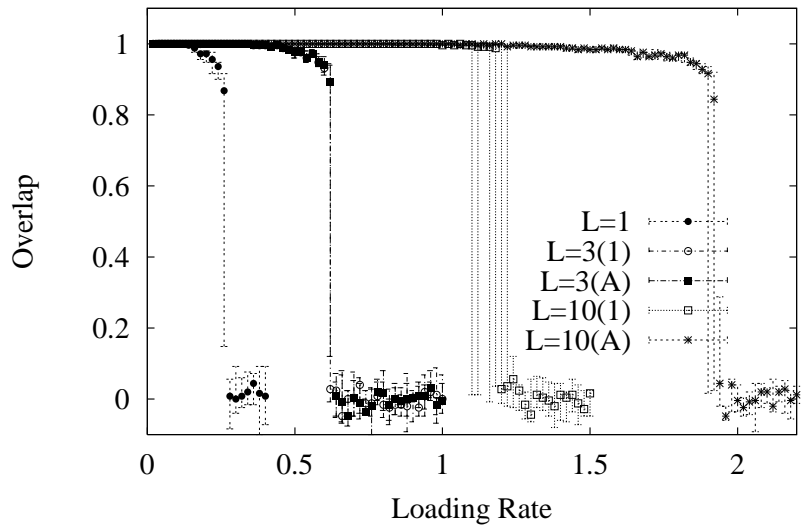}
\caption{Relationship between 
loading rate $\alpha$ and overlap $m$.
``(1)" and ``(A)" indicate one step set initial condition
 and all steps set initial condition, respectively 
(computer simulation with $N=500$).}
\label{fig:Ld1_As}
\end{center}
\end{figure}

Figure \ref{fig:Ld1_At} shows the relationship between 
the loading rates $\alpha$ and the steady state overlaps $m_{\infty}$
obtained by time-dependent 
theoretical calculations
with a sufficient number of steps. 
Figure \ref{fig:Ld1_As} shows the results of computer simulations
under the same conditions as the theoretical calculations.
In each figure, ``(1)'' and ``(A)'' denote the one step set
initial condition and the all steps initial condition, respectively.
In the computer simulations, the number of neurons is $N=500$.
The computer simulations have been carried out
under five conditions :
$L=1$, 
$L=3$ from the one step set initial condition,
$L=3$ from the all steps initial condition,
$L=10$ from the one step set initial condition, and
$L=10$ from the all steps initial condition. 
Eleven simulations have been carried out at various loading rates 
$\alpha$ under each condition.
Here, the initial overlap $m_{{\rm init}}=1$ in 
all cases．Therefore, the all steps set initial condition is
equivalent to the optimum initial condition as described above.
In Figure \ref{fig:Ld1_As}, 
data points $\bullet$ , $\circ$ ,
{\scriptsize $\blacksquare$} , {\scriptsize $\square$} ,
$*$ indicate the medians 
of the 6th largest values in the eleven trials. 
Error bars indicate the third and the ninth 
largest values in the eleven trials.

These figures show that the steady states
obtained by the re-derived theory
are in good agreement with those obtained by computer simulation.
In the case of $L=3$, the difference between the phase 
transition point under the one step set initial condition and 
that under the 
optimum initial condition is small. However, in the case of $L=10$,
the difference is large. 
That is, the influence of the initial 
condition of the network 
increases as the length of delay $L$ increases. 
We note that in the range between the 
phase transition point under the one step set initial condition and
that under the optimum initial condition, 
there is the phenomenon that the attractor
cannot be recalled from the former initial condition,
although the sequence of memory patterns
is certainly stored.

\begin{figure}[htbp]
\begin{center}
\includegraphics[width=0.70\linewidth,keepaspectratio]{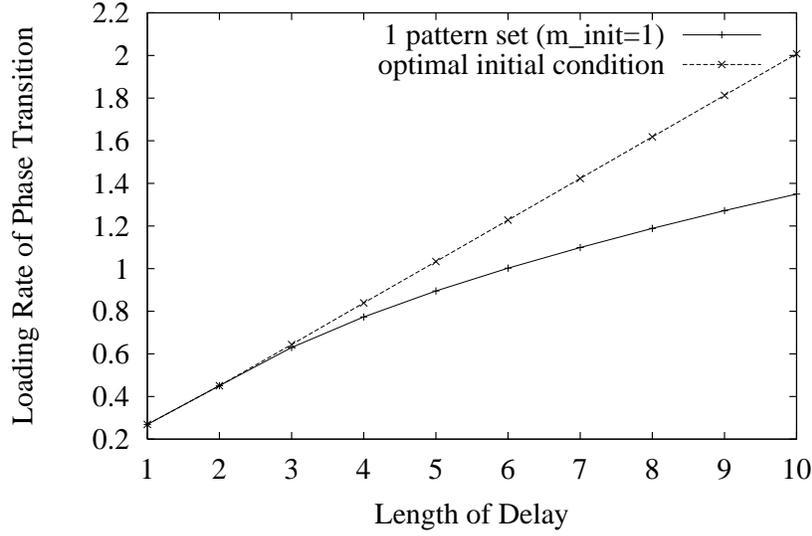}
\caption{Length of delay $L$ and loading rate
of phase transitions. 
Solid line and dashed line show results 
for one step set initial condition
and those for all steps set initial condition,
respectively.}
\label{fig:Ld1_Acapat}
\end{center}
\end{figure}

Figure \ref{fig:Ld1_Acapat} shows the relationship between
the length of delay $L$ and the critical loading rate 
where the phase transition occurs.
This theory (Yanai-Kim theory) needs a computational complexity of
${\cal O}(L^4t)$ to obtain the macrodynamics, 
where $L$ and $t$ are the length of
delay and the time step, respectively.
Therefore, in this method,
it is intractable to investigate the critical loading rate 
in the case of such a large delay $L$.
Here, the results in the cases of $L=1,2,\cdots,10$ are shown.
This figure shows the following. When the initial conditions are
optimum, the relationship between the length of delay $L$ and the
critical loading rate, that is, the 
storage capacity, seems to be almost linear. It is assumed that
when the length of delay $L$ further increases, 
this tendency would continue. 
These characteristics are analyzed 
in the next section.
On the other hand, in the case of the one step set initial condition,
the critical loading rate
is not linear with the length of delay $L$ but saturated.
However, the reason for this phenomenon is 
the lack of initial information. 
We note that
this saturation doesn't imply the essential saturation of the 
storage capacity of the delayed networks.

\section{Macroscopic steady state analysis by discrete
Fourier transformation and discussion}
The Yanai-Kim theory re-derived in the previous section,
that is, the macrodynamical equations obtained by the statistical 
neurodynamics, needs a computational complexity of
${\cal O}(L^4t)$ to obtain the macrodynamics
shown in eqns (\ref{eqn:sigmat2D}) and (\ref{eqn:vD}),
where $L$ and $t$ are the length of
delay and the time step, respectively.
Therefore, in this method,
it is intractable to investigate the critical loading rate 
for a large $L$ limit, that is, the 
asymptotic behavior of the storage capacity in the 
large $L$ limit.
Thus, in this section, the Yanai-Kim theory in a steady state
is considered. After that, we derive the macroscopic steady
state equations of the delayed
network by using the discrete Fourier transformation,
where the computational 
complexity does not formally depend on $L$.
Furthermore, the storage capacity is analytically discussed
for a large $L$ by solving the derived equations numerically.

For simplicity, let us assume $c_l=1,\ l=0,\cdots,L-1$.
In a steady state, $\sigma_t, U_t$ and $m_t$ in 
eqns (\ref{eqn:sigmat2D})-(\ref{eqn:mt1D}) can be expressed
as $\sigma , U$ and $m$, respectively.
In addition, $v_{t-l,t-l'}$ can be expressed as $v_{l-l'}$
because of the parallel symmetry in terms of $v$.
Therefore, modifying eqns (\ref{eqn:sigmat2D}) and (\ref{eqn:vD}), 
we obtain
\begin{eqnarray}
\sigma^2 &=& \sum_{n=1-L}^{L-1}\left(L-|n|\right)v\left(n\right) ,
\label{eqn:sigmasteady}\\
v\left(n\right) &=& \alpha \delta_{n,0} \nonumber \\
     &+& U^2\sum_{i=1-L}^{L-1}\left(L-|i|\right)v\left(n-i\right)
         +\alpha Ud\left(n\right) , \label{eqn:vsteady}\\
d\left(n\right) &=& \left\{
   \begin{array}{ll}
      1,            & |n|=1,2,\cdots,L  \\
      0,            & \mbox{otherwise}
   \end{array}
   \right. ,
\label{eqn:dn}
\end{eqnarray}
where $n=l-l' , i=k-k'$, $v(n)$ denotes $v_{n}$
and $\delta$ is Kronecker's delta.

Using the discrete Fourier transformation, we derive 
the general term of $v(n)$, which is expressed
by the recurrence formula in eqn (\ref{eqn:vsteady}).
Applying the discrete Fourier transformation to
eqns (\ref{eqn:vsteady}) and (\ref{eqn:dn}), we obtain
\begin{eqnarray}
V(r) &=& \alpha+U^2\sum_{i=1-L}^{L-1}(L-|i|)V(r)
e^{-j2\pi \frac{ri}{2T+1}}+\alpha UD(r) ,\label{eqn:Vk}\\
D(r) &=& \sum_{n=-T}^T d(n)e^{-j2\pi \frac{rn}{2T+1}} \nonumber \\
&=& \sum_{n=1}^L 
\left(e^{-j2\pi \frac{rn}{2T+1}}+e^{j2\pi \frac{rn}{2T+1}}\right)
\nonumber \\
&=& 2\sum_{n=1}^L \cos \left(\frac{2\pi rn}{2T+1}\right) ,
\label{eqn:Dk}
\end{eqnarray}
where $V(r)$ and $D(r)$ are the discrete Fourier transformation
of $v(n)$ and $d(n)$, respectively.

Solving eqns (\ref{eqn:Vk}) and (\ref{eqn:Dk}) in terms of $V(r)$,
we obtain
\begin{equation}
V\left(r\right)=\frac{\alpha \left( 1+2U\sum_{n=1}^L\cos
         \left(\frac{2\pi rn}{2T+1}\right)\right)}
         {1-U^2\sum_{i=1-L}^{L-1}(L-|i|)e^{-j2\pi \frac{ri}{2T+1}}} .
\label{eqn:Vk2}
\end{equation}

Since the inverse discrete Fourier transformation of this equation
equals $v\left(n\right)$, we obtain
\begin{equation}
v(n)=\frac{1}{2T+1}\sum_{r=-T}^T V(r)e^{j2\pi \frac{rn}{2T+1}} .
\label{eqn:vn}
\end{equation}

Substituting eqn (\ref{eqn:vn}) into eqn (\ref{eqn:sigmasteady}),
we obtain
\begin{eqnarray}
\sigma^2
&=& \frac{1}{2T+1}\sum_{r=-T}^{T}V\left(r\right)
\sum_{n=1-L}^{L-1}\left(L-|n|\right)e^{j2\pi\frac{rn}{2T+1}} .
\nonumber \\
\label{eqn:sig2}
\end{eqnarray}

Substituting eqn (\ref{eqn:Vk2}) into eqn (\ref{eqn:sig2})
and calculating the equation in the large $T$ limit, we can
express $\sigma^2$ as the form using a simple integral like 
eqn (\ref{eqn:sigma2steady}). 

As a result, we can obtain the steady state equations 
in terms of the macroscopic variables of the network
as eqns (\ref{eqn:sigma2steady})-(\ref{eqn:mm}).
\begin{figure*}[hbt]
\begin{equation}
\sigma^2
= \int_{-\frac{1}{2}}^{\frac{1}{2}}
\frac{\alpha \left[\left(1-U\right)\sin (\pi x) + 
U \sin \left\{\left(2L+1\right)\pi x\right\}\right]
\left[1-\cos (2L\pi x)\right]}
{\sin (\pi x) \left[2\sin^2(\pi x) 
-U^2\left\{1-\cos (2L\pi x)\right\}\right]}
dx \label{eqn:sigma2steady}
\end{equation}
\end{figure*}
\begin{eqnarray}
U &=& \sqrt{\frac{2}{\pi}}\frac{1}{\sigma}
    \exp\left(-\frac{s^2}{2\sigma^2}\right)
    \label{eqn:mUt} \\
s &=& mL \\
m &=& \mbox{erf}\left(\frac{s}{\sqrt{2}\sigma}\right) 
\label{eqn:mm}
\end{eqnarray}

Though the derived macroscopic steady state equations 
include a simple integral, their computational 
complexity does not formally depend on $L$.
Therefore, we can easily perform numerical calculations
for a large $L$.

\begin{figure}[htbp]
\begin{center}
\includegraphics[width=0.70\linewidth,keepaspectratio]{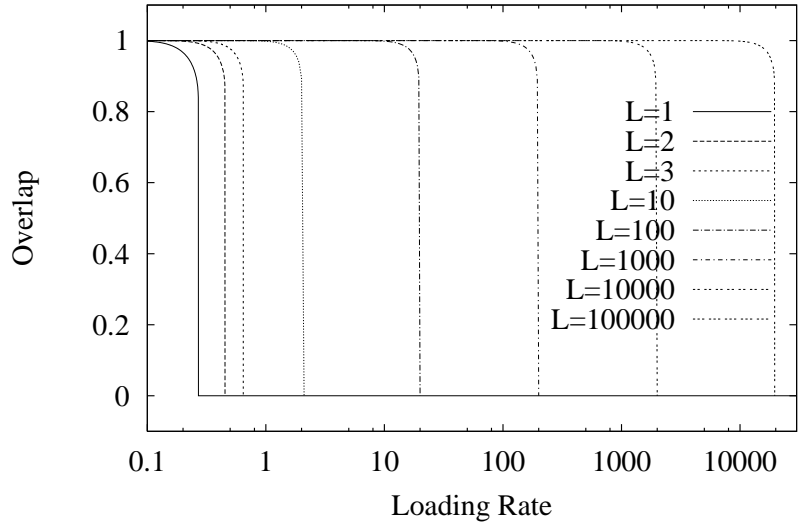}
\caption{Relationship between loading rate $\alpha$ and 
overlap $m$. These lines are obtained by solving
steady state equations numerically.}
\label{fig:Lsnocut}
\end{center}
\end{figure}

Figure \ref{fig:Lsnocut} shows the relationship
between the loading rate $\alpha$ and the overlap $m$,
which is obtained by solving these equations numerically.

Comparing Figure \ref{fig:Ld1_At} and \ref{fig:Lsnocut},
especially considering the case of $L=1,3,10$,
we can see that the overlaps obtained by the macroscopic
steady state equations in this section agree with those 
obtained by the dynamical calculations 
having sufficient time steps 
from the optimum initial conditions in the previous section.
In other words, the phase transition points obtained 
by the macroscopic steady state equations derived in this
section agree with the storage capacity, that is, the phase
transition points under the optimum initial condition.
This fact shows that the solution to the dynamical equations
from the optimum initial condition and the solution to 
the steady state equations support each other.

\begin{figure}[htbp]
\begin{center}
\includegraphics[width=0.70\linewidth,keepaspectratio]{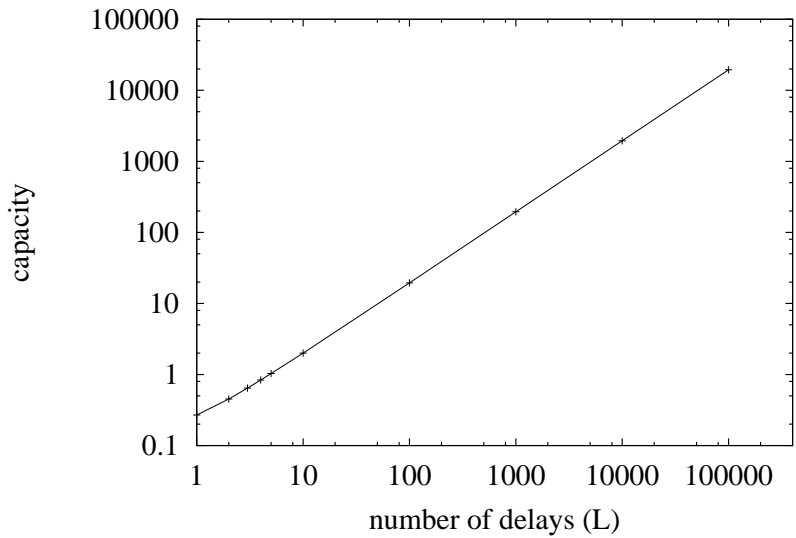}
\caption{Relationship between length of delay $L$ and storage 
capacity $\alpha_C$. This line is obtained by solving the
steady state equations numerically. Storage capacity is $0.195L$
with large $L$ limit.}
\label{fig:nocut}
\end{center}
\end{figure}

Figure \ref{fig:nocut} shows the relationship between
the length of delay $L$ and the storage capacity $\alpha_C$.
From this figure, we can see that the storage capacity 
increases in proportion to the length of delay $L$ with a 
large $L$ limit and the proportion constant is 0.195.
That is, the storage capacity of the delayed network
$\alpha_C$ equals $0.195L$ when the length of delay $L$
is large.
Though the result that the storage capacity of the delayed network
is in proportion to the length of delay $L$ is not nontrivial,
the fact that this result has been proven analytically is
significant. The proportion constant 0.195 is 
the mathematically significant
number as the limit of the delayed network's storage capacity.

\section{Conclusions}
We analyzed sequential 
associative memory models with delayed synapses.
First, we re-derived the Yanai-Kim theory,
which involves
the macrodynamical equations
for networks with serial delay elements.
Since 
their theory needs a computational complexity of ${\cal O}(L^4 t)$
to obtain the macroscopic state at time step $t$
where $L$ is the length of delay,
it is intractable to discuss the macroscopic properties
for a large $L$ limit.
Thus, we derived steady state equations
using the discrete Fourier transformation,
where the complexity does not formally depend on $L$.
We showed that
the storage capacity $\alpha_C$ is in proportion to delay length $L$
with a large $L$ limit,
and the proportion constant is 0.195, i.e., $\alpha_C = 0.195 L$.
These results were supported by computer simulation.

\section*{Appendix A. Derivations of the macrodynamical equations of 
delayed network} \label{sec:A1}
Derivations of the macrodynamical equations of 
delayed network eqns (\ref{eqn:sigmat2D})-(\ref{eqn:mt1D})
discussed in Section \ref{sec:dynamical} are given here.
Using eqns (\ref{eqn:zit}) and (\ref{eqn:mmut2}), we obtain
\begin{eqnarray}
z_i^t &=& z_A+z_B \label{eqn:zitAB} , \\
z_A &=& \sum_{l=0}^{L-1}c_l\sum_{\mu\neq t} \xi_i^{\mu+1}\frac{1}{N}
\sum_j\xi_j^{\mu-l}x_j^{t-l\left(\mu-l\right)} , \label{eqn:zA} \\
z_B &=& \sum_{l=0}^{L-1}c_l\sum_{\mu\neq t} \xi_i^{\mu+1}
U_{t-l}\sum_{l'=0}^{L-1}c_{l'}m_{\mu-l-l'-1}^{t-l-l'-1} ,
\label{eqn:zB}
\end{eqnarray}
where $x_j^{t-l\left(\mu-l\right)}$ is the variable 
obtained by removing
the 
influence of $\xi_j^{\mu-l}$ from $x_j^{t-l}$.
Using eqns (\ref{eqn:zitAB})-(\ref{eqn:zB}),
we obtain
\begin{eqnarray}
E\left[z_i^t\right] &=& 0 , \\
.\raisebox{1ex}{.}. \ \ \ \ \ 
\sigma_t^2 &=& E\left[\left(z_i^t\right)^2\right] \label{eqn:sigmat20}\\
 &=& E\left[z_A^2+z_B^2+2z_Az_B\right] .\label{eqn:sigmat2}
\end{eqnarray}

Transforming $z_A^2,z_B^2,z_Az_B$ with consideration given to 
their correlation, we obtain
\begin{eqnarray}
E\left[z_A^2\right] &=& \alpha \sum_{l=0}^{L-1}c_l^2 ,\label{eqn:A2} \\
E\left[z_B^2\right] &=& \sum_{\mu\neq t}\sum_{l=0}^{L-1}\sum_{l'=0}^{L-1}
\sum_{k=0}^{L-1}\sum_{k'=0}^{L-1}c_lc_{l'}c_kc_{k'} ,\nonumber \\
 & & \times U_{t-l}U_{t-l'}m_{\mu-l-k-1}^{t-l-k-1}m_{\mu-l'-k'-1}^{t-l'-k'-1} 
\label{eqn:B2} \\
E\left[2z_Az_B\right] &=& 
 \alpha\sum_{l=0}^{L-1}\sum_{l'=0}^{L-1}c_lc_{l'} \nonumber \\
& & \times \left(c_{l-l'-1}U_{t-l'}+c_{l'-l-1}U_{t-l}\right) ,
\label{eqn:2AB}
\end{eqnarray}
where
\begin{equation}
v_{t-l,t-l'}=\sum_{\mu\neq t}m_{\mu-l}^{t-l}m_{\mu-l'}^{t-l'} .
\end{equation}

Using eqns (\ref{eqn:sigmat2})-(\ref{eqn:2AB}), we obtain
\begin{eqnarray}
\sigma_t^2 &=& \alpha \sum_{l=0}^{L-1}c_l^2 \nonumber \\
&+& \sum_{l=0}^{L-1}\sum_{l'=0}^{L-1}
\sum_{k=0}^{L-1}\sum_{k'=0}^{L-1}c_lc_{l'}c_kc_{k'} \nonumber \\
 & & \times U_{t-l}U_{t-l'}v_{t-l-k-1,t-l'-k'-1}
\nonumber \\
&+& \alpha\sum_{l=0}^{L-1}\sum_{l'=0}^{L-1}c_lc_{l'}
\left(c_{l-l'-1}U_{t-l'}+c_{l'-l-1}U_{t-l}\right) .
\label{eqn:sigmat22}
\end{eqnarray}

Using eqns (\ref{eqn:zit}) and (\ref{eqn:sigmat20}), we obtain
\begin{equation}
\sigma_t^2 = \sum_{l=0}^{L-1}\sum_{l'=0}^{L-1}c_lc_{l'}v_{t-l,t-l'} .
\label{eqn:sigmat23}
\end{equation}

Comparing eqns (\ref{eqn:sigmat22}) and (\ref{eqn:sigmat23}) as
identical equations regarding $c_lc_{l'}$,
we obtain
\begin{eqnarray}
v_{t-l,t-l'}
&=& \alpha \delta_{l,l'} \nonumber \\
&+& U_{t-l}U_{t-l'} \nonumber \\
& & \times 
    \sum_{k=0}^{L-1}\sum_{k'=0}^{L-1}c_k c_{k'}v_{t-l-k-1,t-l'-k'-1}
    \nonumber \\
&+& \alpha \left(c_{l-l'-1}U_{t-l'} + c_{l'-l-1}U_{t-l}\right) ,
\label{eqn:v}
\end{eqnarray}
where $\delta$ is Kronecker's delta.
Using eqn (\ref{eqn:Ut}), we obtain
\begin{eqnarray}
U_t
&=& \frac{1}{N}\sum_{i=1}^N 
    F'\left(\sum_{l=0}^{L-1}\sum_{j=1}^N\frac{c_l}{N}\right. 
    \nonumber \\
& & \hspace{10mm} \times \left.\sum_{\nu \neq \mu-l-1}\xi_i^{\nu+1+l}
    \xi_j^\nu x_j^{t-l-1}\right) \\
&=& E\left[F'\left(u^{t\left(\mu\right)}\right)\right] \\
&=& E\left[F'\left(u^t\right)\right] \\
&=& \int\frac{dz}{\sqrt{2\pi}}e^{-\frac{z^2}{2}}\ll F'\left(u^t\right)\gg \\
&=& \frac{1}{\sigma}\int\frac{dz}{\sqrt{2\pi}}e^{-\frac{z^2}{2}}z
\ll F\left(u^t\right)\gg \\
&=& \sqrt{\frac{2}{\pi}} \frac{1}{\sigma_{t-1}}
    \exp\left(-\frac{\left(s^{t-1}\right)^2}{2\sigma_{t-1}^2}\right) ,
    \label{eqn:Ut2}
\end{eqnarray}
where $u^{t\left(\mu\right)}$ is the variable 
obtained by removing the
influence of $\mbox{\boldmath $\xi$}^\mu$ from $u^{t}$.
$\ll\cdot\gg$ stands for the average over pattern 
$\mbox{\boldmath $\xi$}$.

As a result, we can obtain the macrodynamical equations
for overlap $m$, that is,
eqns (\ref{eqn:sigmat2D})-(\ref{eqn:mt1D}).

%

\section*{References}

\begin{description}

\item
Amari, S., \& Maginu, K. (1988).
\newblock Statistical neurodynamics of associative memory.
\newblock {\em Neural Networks, 1}, 63--73.

\item
Amari, S. (1988).
\newblock Statistical neurodynamics of various versions of
correlation associative memory.
\newblock {\em Proceedings of IEEE International 
Conference on Neural Networks, 1}, 633--640.

\item
Amit, D.J., Gutfreund, H., \& Sompolinsky, H. (1985a).
\newblock Spin-glass model of neural networks.
\newblock {\em Physical Review A, 32}, 1007--1018.

\item
Amit, D.J., Gutfreund, H., \& Sompolinsky, H. (1985b).
\newblock Storing infinite numbers of patterns in a spin-glass 
model of neural networks.
\newblock {\em Physical Review Letters, 55}, 1530--1533.

\item
D\"{u}ring, A., Coolen, A.C.C., \& Sherrington, D. (1998).
\newblock Phase diagram and storage capacity of sequence processing
neural networks.
\newblock {\em Journal of Physics A: Mathematical and General, 31},
8607--8621.

\item
Fukushima, K. (1973).
\newblock A model of associative memory in the brain.
\newblock {\em Kybernetik, 12}, 58--63.

\item
Hopfield, J.J. (1982).
\newblock Neural networks and physical systems with emergent collective
computational abilities.
\newblock {\em Proceedings of National Academy of Sciences, 79}, 2554--2558.

\item
Kawamura, M., \& Okada, M. (2002).
\newblock Transient dynamics for sequence processing neural
networks.
\newblock {\em Journal of Physics A: Mathematical and General, 35}, 253--266.

\item
Miyoshi, S., \& Nakayama, K. (1995).
\newblock A recurrent neural network with serial delay elements for 
memorizing limit cycles.
\newblock {\em Proceedings of IEEE International 
Conference on Neural Networks}, 1955--1960. 

\item
Miyoshi, S., \& Okada, M. (2000).
\newblock A theory of syn-fire chain model.
\newblock {\em Transactions of the Institute of Electronics,
Information and Communication Engineeres, J83-A} (11), 1330-1332.
in Japanese.

\item
Miyoshi, S., \& Okada, M. (2002).
\newblock Associative memory by neural networks with delays and pruning.
\newblock {\em Transactions of the Institute of Electronics,
Information and Communication Engineeres, J85-A} (1), 124-133.
in Japanese. 

\item
Okada, M. (1995).
\newblock A hierarchy of macrodynamical equations for
associative memory.
\newblock {\em Neural Networks, 8} (6), 833--838.

\item
Okada, M. (1996).
\newblock Notions of associative memory and sparse coding.
\newblock {\em Neural Networks, 9} (8), 1429--1458.

%

\item
Sherrington, D., \& Kirkpatrick, S. (1975).
\newblock Solvable model of a spin-glass.
\newblock {\em Physical Review Letters, 35}, 1792--1796.

\item
Shiino, M., \& Fukai, T. (1992).
\newblock Self-consistent signal-to-noise analysis and
its application to analogue neural networks with asymmetric connections.
\newblock {\em Journal of Physics A: Mathematical and General, 25},
L375--L381.
%
%

\item
Yanai, H.-F., \& Kim, E.S. (1993).
\newblock Dynamics of neural nets with delay-synapses.
\newblock {\em Technical report of the Institute of Electronics,
Information and Communication Engineeres}, NC92-116, 167--174.
in Japanese.

\end{description}

\end{document}